\documentclass[review]{elsarticle}

\usepackage{hyperref,makecell,amssymb,amsmath}
\usepackage{amstext,amsfonts}
\usepackage{graphicx,subfigure}
\usepackage{caption}
\usepackage{subfigure}
\usepackage{color}
\usepackage{pdfpages}

\journal{arXiv:1804.04499}

\usepackage{draftwatermark}

\begin{document}
\captionsetup[figure]{labelfont={bf},name={Fig.},labelsep=period}
\begin{frontmatter}

\title{Energy Response of GECAM Gamma-Ray Detector Based on LaBr$_{3}$:Ce and SiPM Array}

\author[]{Dali Zhang\corref{mycorrespondingauthor}}
\cortext[mycorrespondingauthor]{Corresponding author}
\ead{zhangdl@ihep.ac.cn}

\author[]{Xinqiao Li}
\author[]{Shaolin Xiong}
\author[]{Huanyu Wang}
\author[]{Fan Zhang}

\author[]{Wenxi Peng}

\author[]{Yanguo Li}
\author[]{Xilei Sun}
\author[]{Zhenghua An}
\author[]{Yanbing Xu}
\author[]{Yue Zhu}

\address{Key Laboratory of Particle Astrophysics, Institute of High Energy Physics, Chinese Academy of Sciences, Beijing, China}

\begin{abstract}
The Gravitational wave high-energy Electromagnetic Counterpart All-sky Monitor (GECAM) , composed of two small satellites, is a new mission to monitor the Gamma-Ray Bursts (GRBs) coincident with gravitational wave(GW) events with a FOV of 100\% all-sky. Each GECAM satellite detects and localizes 6 keV--5 MeV GRBs via 25 compact and novel Gamma-Ray Detectors (GRDs). Each GRD module is comprised of a LaBr$_{3}$:Ce scintillator, SiPM array and preamplifier. A large dynamic range of GRD is achieved by the high gain and low gain channels of the preamplifier. The performance of GRD prototype was evaluated using radioactive sources in the range of 5.9--1332.5 keV. A energy resolution of 5.3\% at 662 keV was determined from the $^{137}$Cs pulse height spectra and it meets the GECAM requirement ($<$ 8\% at 662 keV). Energy to channel conversion was evaluated and a nonlinearity correction was performed to reduce the residuals ($<$ 1.5\%). A Geant4 based in-flight environment background simulation and measured GRD LaBr$_{3}$:Ce intrinsic activity were used to evaluate the capability of in-flight calibration. These results show the GRD based on SiPM array can use the LaBr$_{3}$:Ce intrinsic activity for in-flight calibration.
\end{abstract}


\end{frontmatter}


\section{Introduction}
\par On August 17, 2017, the Advanced LIGO and Advanced Virgo jointly detected gravitational waves (GW170817) originating from a binary neutron star coalescence\cite{GW170817}. The Gamma-ray Burst Monitor (GBM) onboard Fermi\cite{GBM} detected a short Gamma-Ray Burst (GRB) with a time delay of $\backsim$1.7 s with respect to the merger time. A global multi-wavelength observation campaign was started in response to the LIGO-Virgo and GBM detections to monitor the electromagnetic counterpart of gravitational waves\cite{MultiMes}. Many gamma-ray telescopes for monitoring GRBs, including BurstCube\cite{BurstCube} and MERGR\cite{MERGR}, have been proposed given the critical importance of detecting GRBs associated with GW sources.

\par
\begin{figure}
  \centering
  \includegraphics[width=5.2 cm]{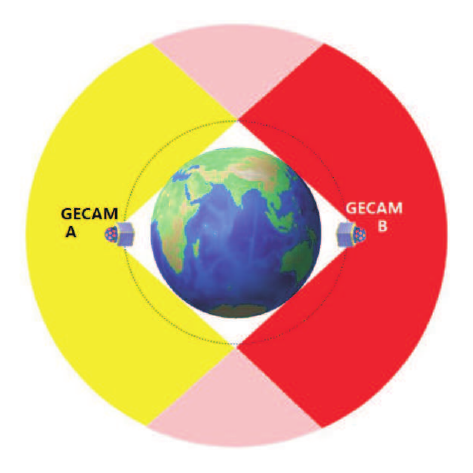}
  \includegraphics[width=6 cm]{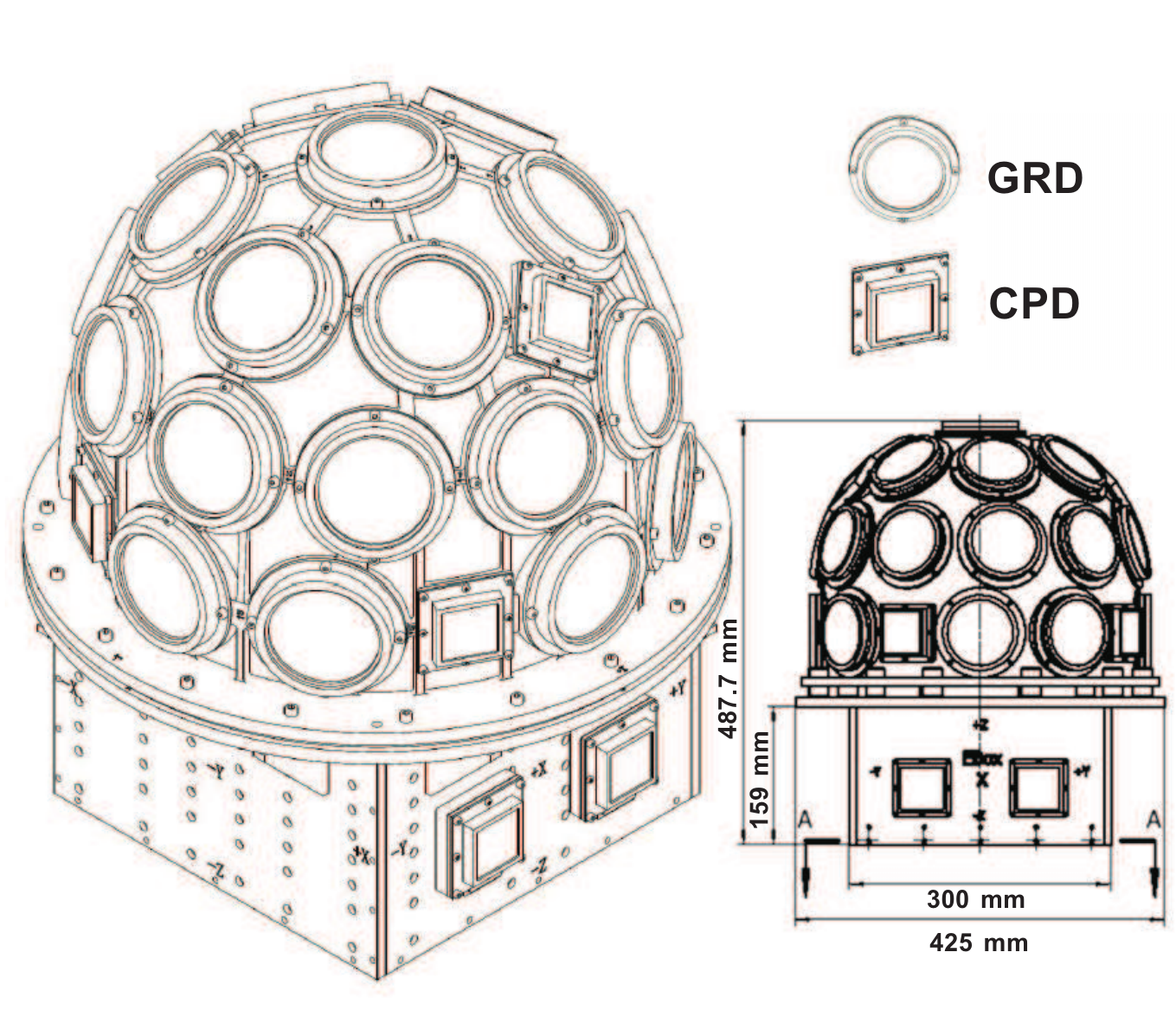}
    \caption{Left:GECAM consists of two small satellites with opposite orbital phase; Right: Each GECAM satellite consists of 25 cylindrical GRDs and 8 square CPDs. {\color{black} The size of the payload is marked in the picture.}}\label{GECAMstructure}
\end{figure}

 GECAM is a new Chinese mission for monitoring the GW associated GRBs which was proposed in the March of 2016, just after the announcement of the first discovery of GW source by LIGO. GECAM’s primary science objective is to detect and localize X-ray and gamma-ray emission from 6 keV to 5 MeV (the requirement is from 8 keV to 2 MeV) of the GW events detected by LIGO and Virgo. Unlike most of other gamma-ray missions, GECAM consists of two small satellites (Fig.\ref{GECAMstructure} left) which operate in the same low earth orbit ($\backsim$600 km) but with opposite orbital phase.{\color{black} The designed sensitivity of GECAM is 2$\times$10$^{-8}$ erg/cm/s and angular resolution will be better than 1 degree.} Each GECAM satellite (Fig.\ref{GECAMstructure} right) consists of 25 Gamma-Ray Detector (GRD) and 8 Charged Particle Detector (CPD). {\color{black}The estimated total power consumption is 31 W and weight is 52 kg}. GRDs and CPDs are used to detect gamma-rays and charged particles respectively. By using 25 GRDs with different orientation, each satellite will monitor all-sky un-occulted by the Earth, resulting in that GECAM features a 100\% all-sky FOV. Localization can be reconstructed by relative count rates in 25 GRDs\cite{GBMLocate}. GECAM is planned to launch in 2020 and it will be a very important high energy telescope in the multi-messenger and multi-wavelength era.

\par The innovation of the GRD design is the application of large-volume LaBr$_{3}$:Ce ({\color{black}3 inches in diameter and 15 mm thickness}) and the SiPM array. The LaBr$_{3}$:Ce crystal is one of the best scintillators available, as it features higher light yield and better energy resolution than conventional crystals such as NaI or CsI\cite{LaBrlight}. Another important merit of LaBr$_{3}$:Ce is the stability of its light output \cite{LaBrtemp} during temperature fluctuation in space. {\color{black}Some space payloads use LaBr$_{3}$:Ce crystal to detect GRBs, such as the CGBM onboard the CALET mission\cite{CALET} and Chang'E-2 mission in China\cite{Chang'E2}; The Gamma-Ray Monitor of the proposed GRIPS mission also select LaBr$_{3}$ as calorimeter\cite{GRIPS}.} SiPM arrays are semiconductor devices applied in various fields and common to cutting-edge experiments in astroparticle physics. Its advantages include low bias-voltage (several tens of volts), high detection efficiency, insensitivity to magnetic fields, and compact size. The auto-gain control detector of HXMT\cite{HXMTSiPM}, for example, is equipped with SiPM which measures the count rate of the calibration radioactive source rather than the energy spectra. Some preliminary test results of the GECAM GRD were discussed\cite{GEACAMPreTest}.

\par In this study, we discussed the energy response of a GECAM GRD prototype using a set of radioactive sources. The energy resolution and energy to channel conversion of the GRD detector are presented. The in-flight calibration capability using the LaBr$_{3}$:Ce intrinsic activity is also discussed.

\section{Experimental setup}
\par
\begin{figure}[htbp]
\centering
\includegraphics[width=6.5 cm]{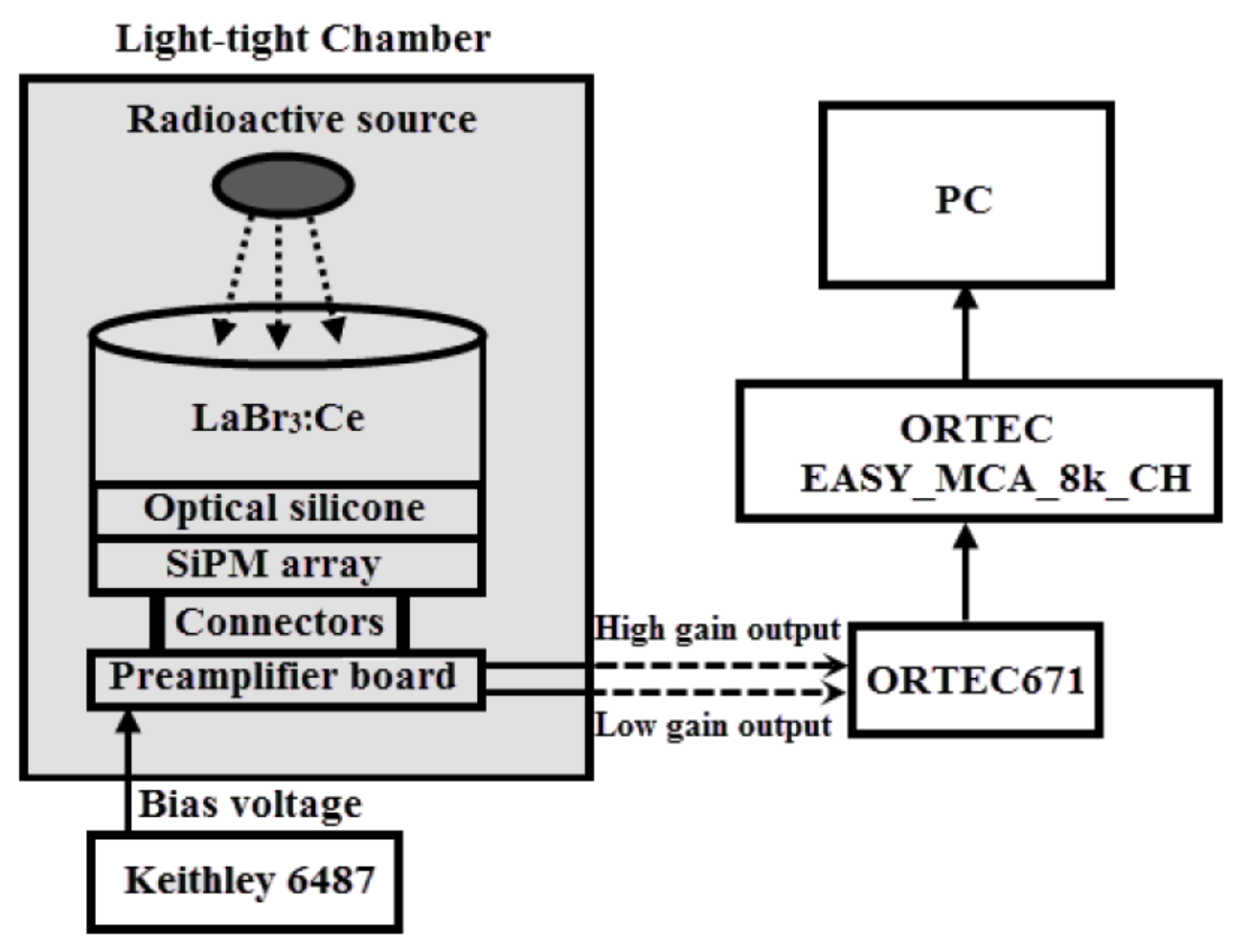}
\includegraphics[width=5.5 cm]{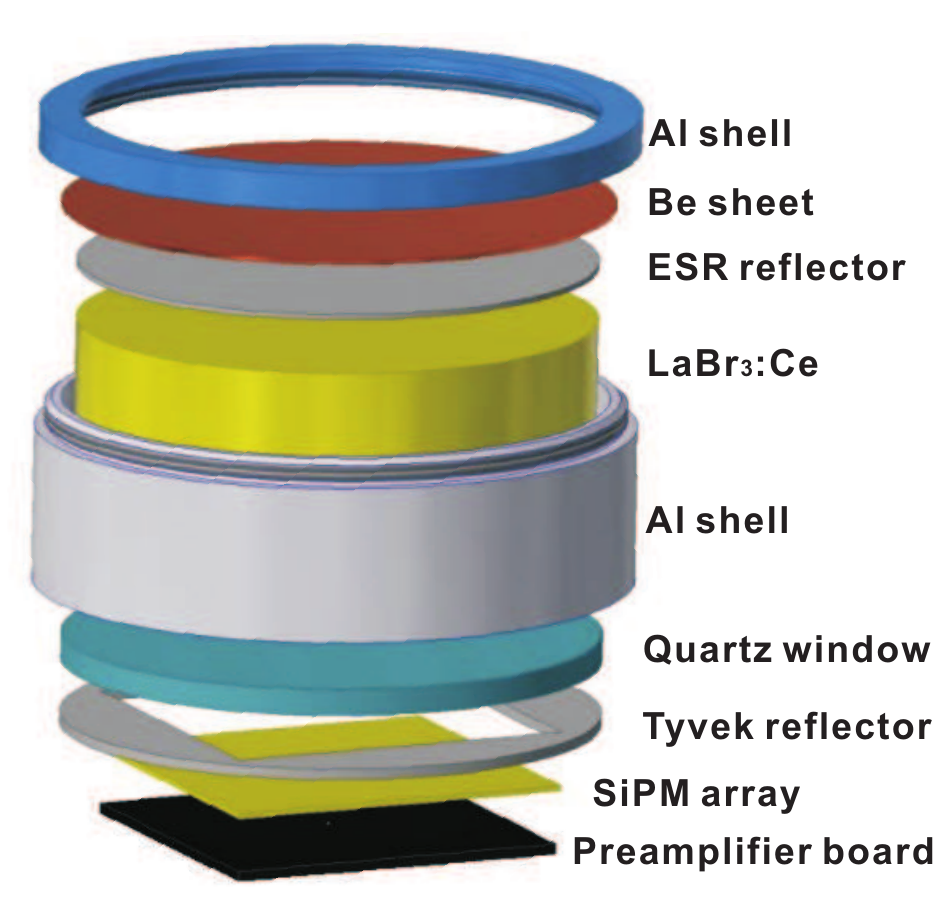}
\caption{\label{Expsetup} Left: Experimental setup; {\color{black}Right: GRD module structure.}}
\end{figure}

 We used a set of radioactive sources ($^{55}$Fe, $^{241}$Am, $^{137}$Cs, $^{60}$Co) to evaluate the performance of the GRD module. These tests cover a major part of the GRD detection energy range. The experimental setup is shown in Fig.\ref{Expsetup} (left). The GRD was placed in an aluminum box to shield it from light and electromagnetic noise. The LaBr$_{3}$:Ce scintillator crystal was supplied by Saint-Gobain. The diameter of the cylindrical crystal is 76.2 mm (3 inches) and its thickness is 15 mm. The crystal was packed in a housing to prevent hygroscopic effects. The entrance window is comprised of a 0.22 mm-thick Be sheet; {\color{black} The ESR reflector was placed behind the Be window and rest of the crystal is wrapped with PTFE to improve the light collection (Fig.\ref{Expsetup} right).} The light output window is equipped with a quartz window for SiPM array coupling through optical silicone (Fig.\ref{Expsetup} right). The 8$\times$8 SiPM array (ARRAYC-60035-64P-PCB) is provided by SensL. The total size of the SiPM array is 50.44 $\times$50.44 mm$^{2}$; it operates at a bias voltage of 28 V, provided by the Keithley 6487 Picoammeter/Voltage source. The output signal of SiPM array was amplified by a homemade preamplifier that has two channels: high gain and low gain. The outputs of preamplifier are shaped by an Ortec 671 spectroscopic amplifier. The high gain or low gain channel can be chosen by changing the connection with Ortec 671. The Ortec 671 output was digitalized by Ortec {\color{black}EASY-MCA-8k-CH} multichannel analyzer (MCA), then the data was transmitted to the PC. The gain of Ortec 671 was set to 10 and the shaping time to 0.5 $\mu$s in Guassian shaping mode.

\par
\begin{figure}
  \centering
  \includegraphics[width=12 cm]{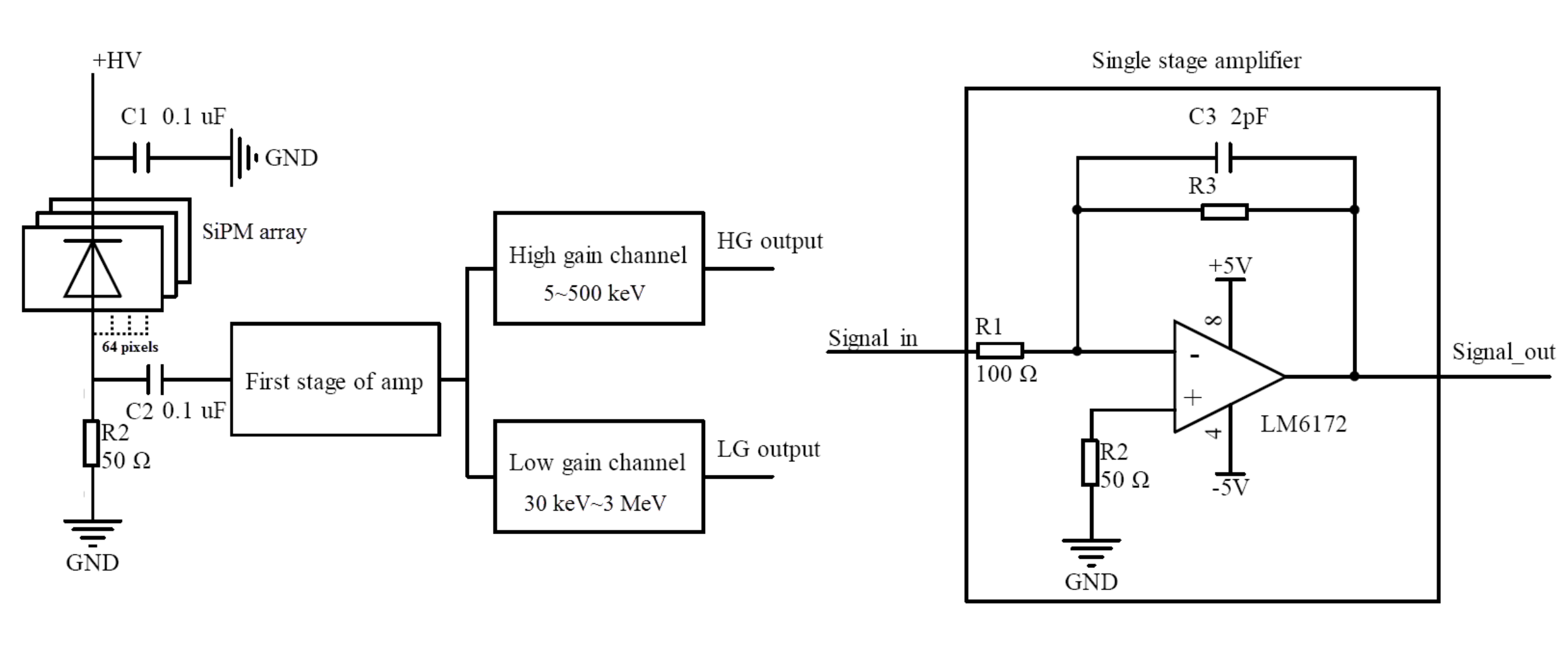}
  \caption{Left: Block scheme of SiPM array and preamplifier. The quench resistor of the SiPM array is 50 $\Omega$ and the preamplifier is capacitively coupled to the SiPM standard output anode with a coupling capacitance of 0.1 $\mu$F.; Right: One-stage preamplifier.}\label{Preamp}
\end{figure}

 To achieve a large dynamic range of GRD, we designed two channels for the preamplifier: high gain and low gain. The SiPM array signal is amplified at the first stage and then fanned out to the high gain stage and low gain stage (Fig.\ref{Preamp} left). The dynamic ranges of high gain and low gain channels are 5--500 keV and 30 keV--3 MeV, respectively. The GRD preamplifier operates in space, so space-qualified devices must be chosen to ensure stable functioning onboard. {\color{black}The preamplifier was developed based on the commercial type of LM6172. This OP amplifier has the space qualified type (LM6172QML) and it is widely used in the space projects in our institute (IHEP).} LM6172 has a high slew rate (3000 V/$\mu$s) and relatively wide unity-gain bandwidth (100 MHz). As shown in Fig.\ref{Preamp} (right), each amplifier stage is implemented as a standard voltage feedback topology and a feedback capacitor (C3) is applied to avoid peaking or oscillation. The gain of each stage can be adjusted by changing the feedback resistor (R3) value. The output of the preamplifier is a positive going pulse with rise time around 400 ns and a fall time around 2 $\mu$s. The power of the preamplifier is $\pm$5 V and the power dissipation of each GRD preamplifier is 100 mW. {\color{black}This preamplifier will have some improvements according to experiment results in the following development phase.}

\section{Results and discussion}
\subsection{GRD LaBr$_{3}$:Ce intrinsic activity measurement}
\par
\begin{figure}
  \centering
  \includegraphics[width=8 cm]{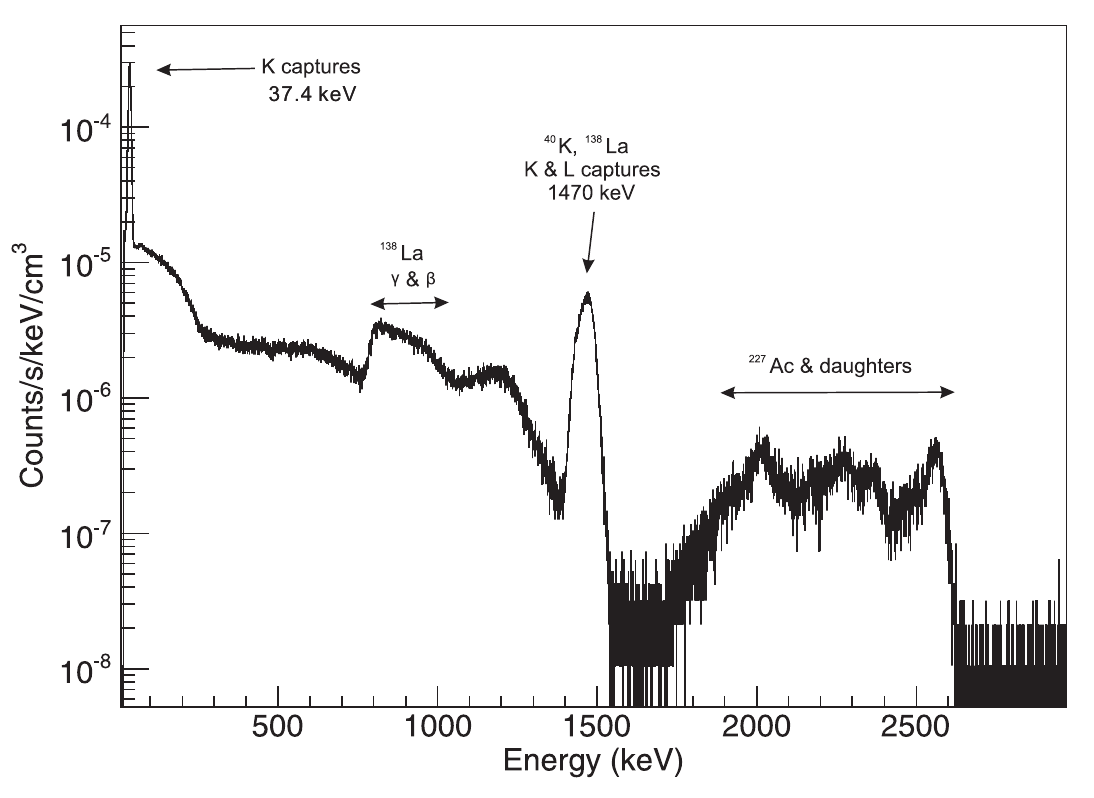}
  \caption{Tested intrinsic activity spectra of the GRD module in energy range from 16 keV to 3 MeV. Activities are given in counts s$^{-1}$ keV$^{-1}$ cm$^{-3}$, equivalent to Bq keV$^{-1}$ cm$^{-3}$.}\label{Labrbackground}
\end{figure}

 The intrinsic activity of LaBr$_{3}$:Ce has two origins: contamination due to the radioactive isotope $^{227}$Ac and its daughters, and the presence of radioactive isotope $^{138}$La\cite{LaBrbackground}. This may be a trouble for low count rate experiments, but it does not affect GECAM measurement because the in-flight background of GECAM is dominated by cosmic x-ray background. On the other hand, the intrinsic activity can also be used for the GRD in-flight calibration. To measure the GRD intrinsic activity, GRD was placed in a 5 cm thick lead chamber to shield the laboratory background. Fig.\ref{Labrbackground} shows the  intrinsic activity spectra in energy range from 16 keV to 3 MeV in 5 hours. The average count rate is 93 counts/s, corresponding to 0.545 counts/s/cm$^{3}$. The 5.6 keV full energy peak of LaBr$_{3}$:Ce intrinsic activity is submerged by the SiPM array dark noise and the tested intrinsic activity spectra only shows the energy range of 16 keV--3 MeV. {\color{black} Because of the non-proportional response of LaBr$_{3}$:Ce scintillation, the measured energy of 37.4 keV from $^{138}$La energy is only 35.5 keV}\cite{LaBrbackground}; The peak around 1.47 MeV, from $^{138}$Ba and $^{40}$K (in the environment), centers at 1470 keV. The three peaks around 1.7--2.8 MeV originate from the alpha decay of $^{227}$Ac and its daughters\cite{LaBrbackground}.

\subsection{Measurements with radioactive sources and energy resolution}
\par We used a set of radioactive sources to characterize the energy response of the GRD in the 5.9--1173.2 keV range, i.e., the majority of the GRD energy range. The SiPM gain is affected by temperature fluctuations. In order to stabilize the SiPM gain, the GRD is placed in an environmental chamber (20 $^{\circ}$C). The full energy peaks of the measured radioactive principle lines were fitted with Gaussian functions and FWHM energy resolutions were extracted.
\par
\begin{figure}[htbp]
\centering
\includegraphics[width=6cm]{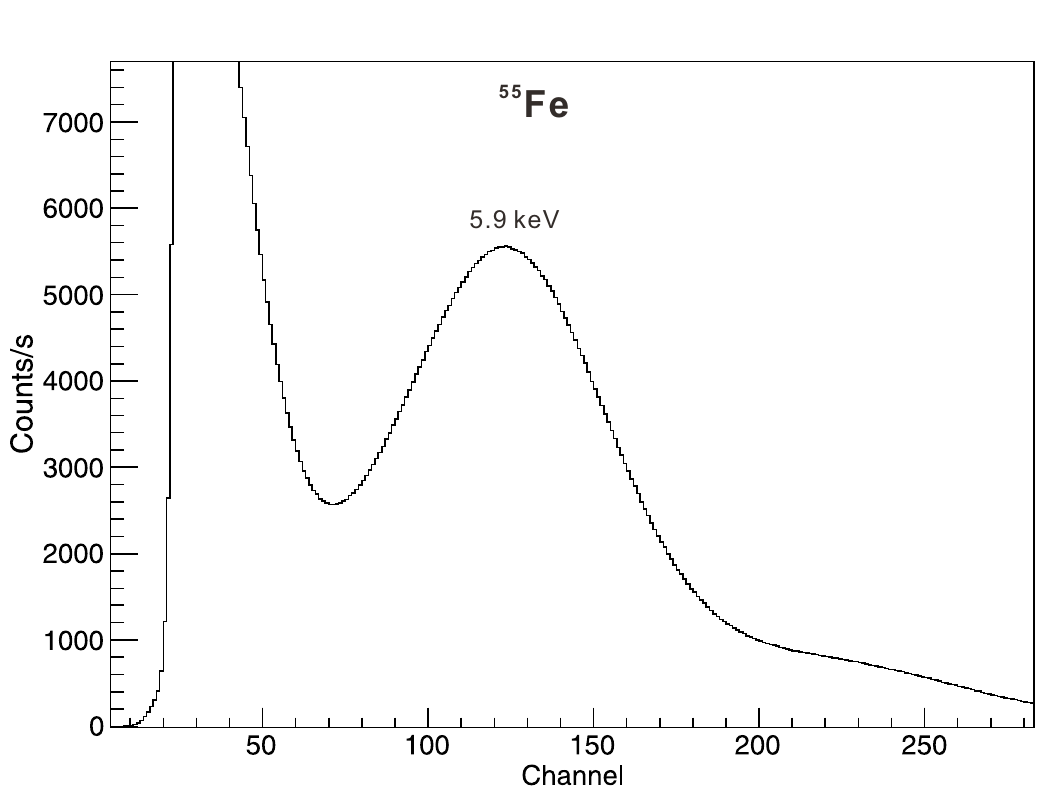}
\includegraphics[width=6cm]{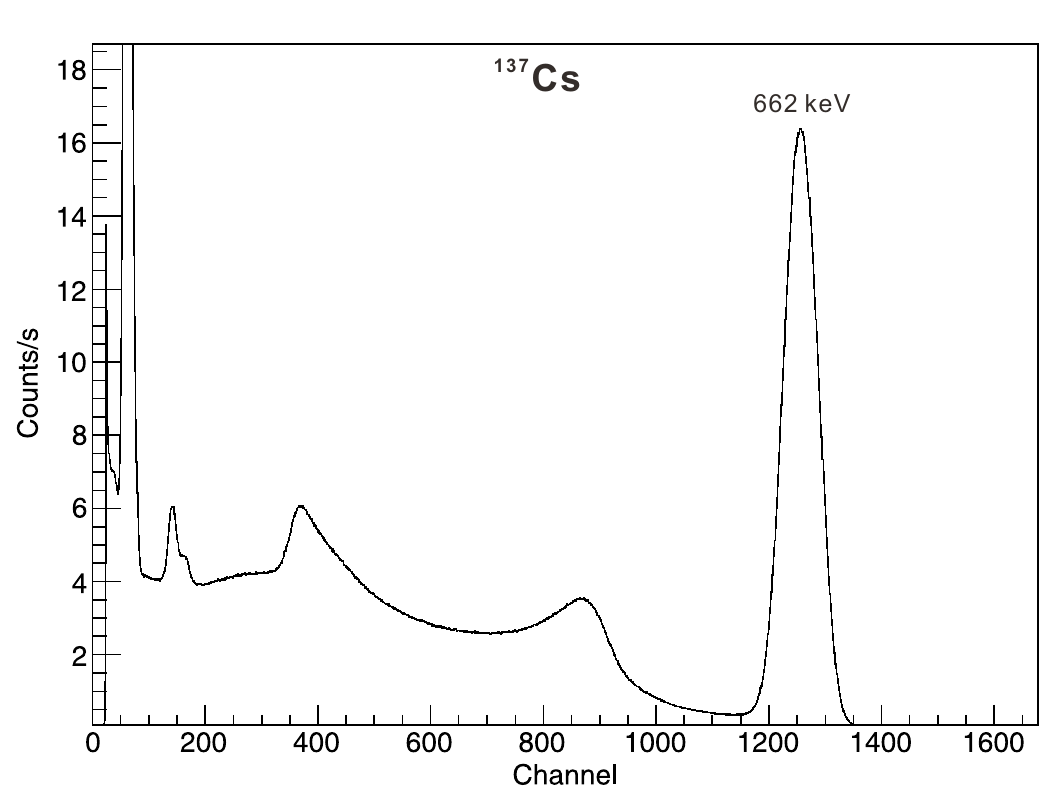}
\caption{\label{Spectra} Measured pulse height histograms. Left: Measured pulse height histogram of $^{55}$Fe with high gain channel. The energy resolution is 65.6\% at 5.9 keV. Right: The pulse height histogram of $^{137}$Cs with low gain channel. The energy resolution is 5.3\% at 662 keV.}
\end{figure}

 In the tests with $^{55}$Fe and $^{241}$Am, the sources were placed directly above the Be window of GRD and high gain channel (5--500 keV) was used. Fig.\ref{Spectra} (left) shows the pulse height histograms of $^{55}$Fe. The gaussian fit shows energy resolution (FWHM) is 65.6\% at 5.9 keV. In the tests with $^{133}$Ba , $^{137}$Cs and $^{60}$Co, the sources were collimated by lead bricks and the low gain channel (30 keV--3 MeV) was used. Fig.\ref{Spectra} (right) shows the $^{137}$Cs pulse height histogram. The gaussian fit shows an energy resolution of 5.3\% at 662 KeV and it meets the GECAM energy resolution requirements ($<$ 8\% at 662 keV). The complete list of the sources used together with their respective gamma-ray and X-ray emission lines and FWHM energy resolution are given in Table \ref{tab1}.

\par
\begin{table}
\caption{Studied X-ray and gamma-ray lines with FWHM resolution for the radioactive test.}\label{1}
\centering
\begin{tabular}{ccc}
\hline
Radioactive sources& Line energy& Energy resolution\\
                   & (keV)      &(\%)              \\
\hline
$^{55}$Fe K$\alpha$ X-ray	&5.9	&65.6	\\
$\gamma$$^{133}$Ba	        &30.85	&23.89	\\
$\gamma$$^{241}$Am	        &59.5	&13.2	\\
$\gamma$$^{133}$Ba	        &81	    &12.64	\\
$\gamma$$^{133}$Ba	        &276.4	&6.65	\\
$\gamma$$^{133}$Ba	        &302.9	&5.87	\\
$\gamma$$^{133}$Ba	        &356	&4.99	\\
$\gamma$$^{133}$Ba	        &383.8	&5.41	\\
$\gamma$$^{137}$Cs	        &662	&5.3	\\
$\gamma$$^{60}$Co	        &1173.2	&4.2	\\
$\gamma$$^{60}$Co	        &1332.5	&3.41	\\
\hline
\end{tabular}
\label{tab1}
\end{table}

\par
\begin{figure}[htbp]
\centering
\includegraphics[width=10cm]{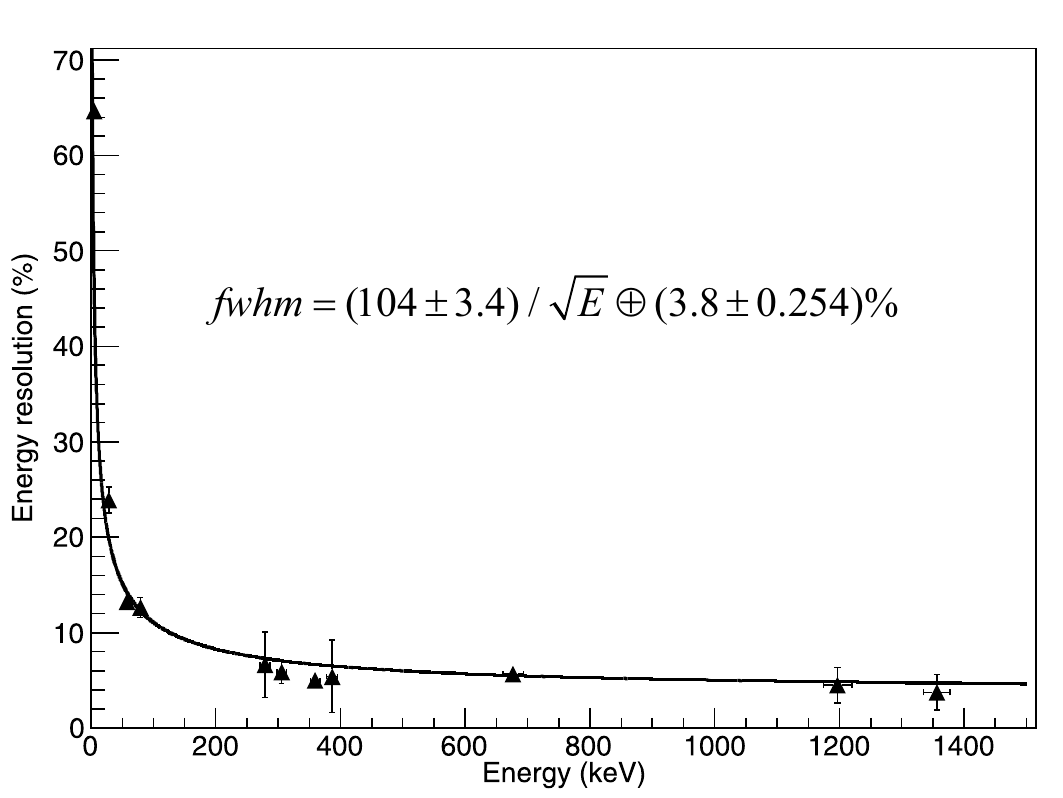}
\caption{\label{resolution} Energy resolution for GRD module. In the low energy range, resolution is mainly affected by statistical uncertainty; in the higher energy range, resolution is mainly affected by GRD non-idealities.}
\end{figure}

 We have evaluated the energy resolution versus energy. The results in Table \ref{tab1} are plotted in Fig.\ref{resolution}. The energy resolution is affected by the uncertainty of the Poisson statistics, which is proportional to E$^{-0.5}$. The data are fitted by:

\begin{equation}
FWHM (\%)=a/\sqrt{E}\oplus b\label{Res_Fit}
\end{equation}

\par where the constant \emph{a} is determined by the light output of the crystal and photo detection efficiency (PDE) of the coupled devices\cite{resolution_fit}; Constant \emph{b} represents the
whole contribution from GRD non-idealities. In the higher energy range, constant \emph{b} becomes more responsible for the energy resolution. Other contributions to the energy resolution such as the non-proportional response of inorganic scintillators\cite{NPR} are not added to the curve because they require more detailed measurements via X-ray calibration facilities in the low energy range (5--300 keV). The results presented here are an approximate evaluation of the GRD energy resolution.

\par The fitting results in Fig.\ref{resolution} show the energy resolution is not as good as those obtained with the PMT\cite{LaBr_PMT}\cite{LaBr_PMT_3D} {\color{black}or SiPM array\cite{LaBr_SiPM}} due to the altogether lower PDE and worse light collection of the GRD module. Although the tested SiPM pixel, with a voltage supply of 28 V at the 420 nm wavelength, has a higher quantum efficiency(40\%) than PMT, the dead area between the pixels of the array results in an overall decrease in PDE; The squared SiPM array covers only 50.5\% of light output window and it results in relatively poor energy resolution in the higher energy range. However, the current results of energy resolution at 662 keV with GRD module falls well within the requirements of GECAM (8\% at 662 KeV).
\par
\begin{figure}[htbp]
\centering
\includegraphics[width=10cm]{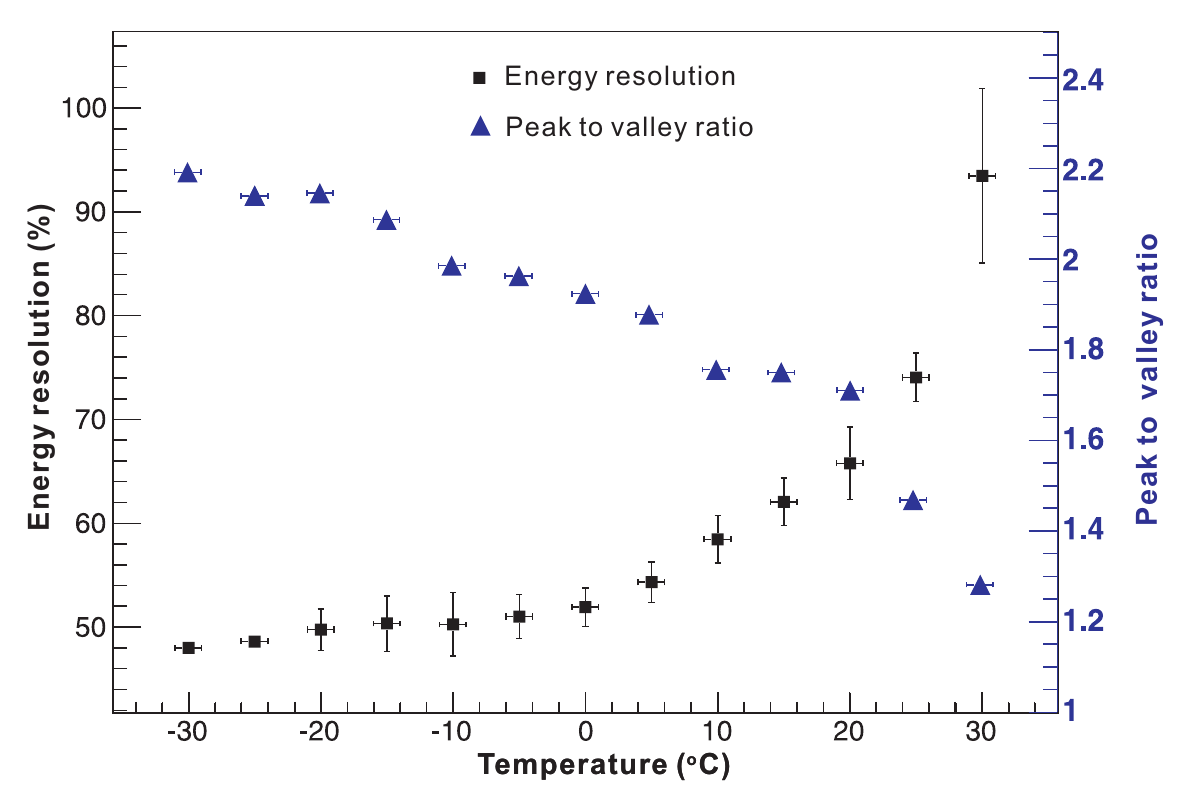}
\caption{\label{Fe55_temp} {\color{black} The temperature behavior of GRD at 5.9 keV  x-ray under -30 $^{\circ}$C to 30 $^{\circ}$C. } }
\end{figure}

 {\color{black} In GECAM thermal design, the temperature variation of LaBr$_{3}$ crystal and SiPM array are -30 $^{\circ}$C to 20 $^{\circ}$C and -30 $^{\circ}$C to -10 $^{\circ}$C, respectively; The temperature change rate will be within 5 $^{\circ}$C/hour. We have prepared a thermostatic chamber in laboratory and measured the behavior of the GRD module under -30 $^{\circ}$C to 30 $^{\circ}$C. The GRD temperature behavior at 5.9 keV is shown in Fig.\ref{Fe55_temp} using $^{55}$Fe radioactive source. As temperature decreases, the energy resolution and peak to valley ratio are better. The energy resolution are 93.4\% at 30 $^{\circ}$C and 44.9\% at -30 $^{\circ}$C. The peak to valley ratio are 1.278 at 30 $^{\circ}$C and 2.198 at -30 $^{\circ}$C).}

\subsection{Energy to channel conversion and nonlinearity correction}
\par
\begin{figure}[htbp]
\centering
\includegraphics[width=5.6cm]{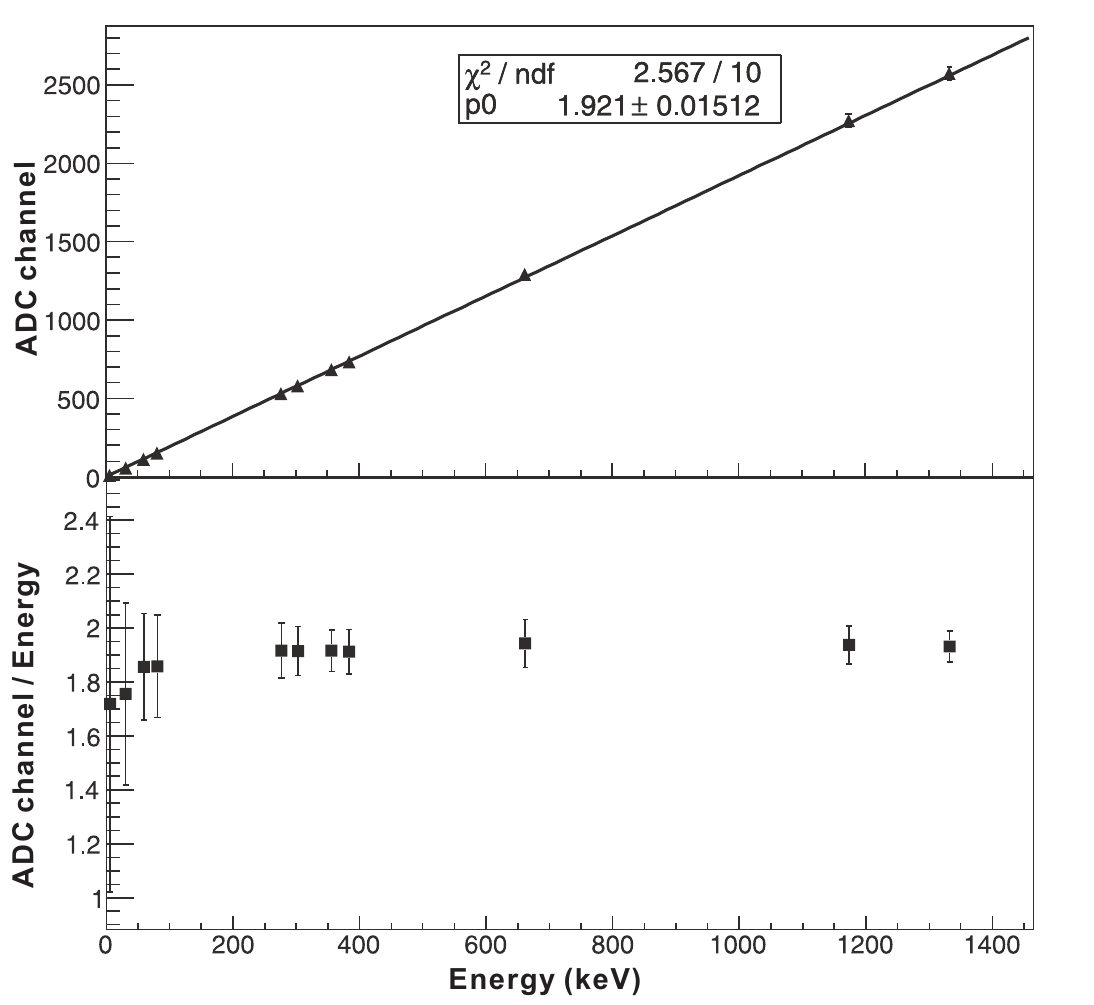}
\includegraphics[width=6.4cm]{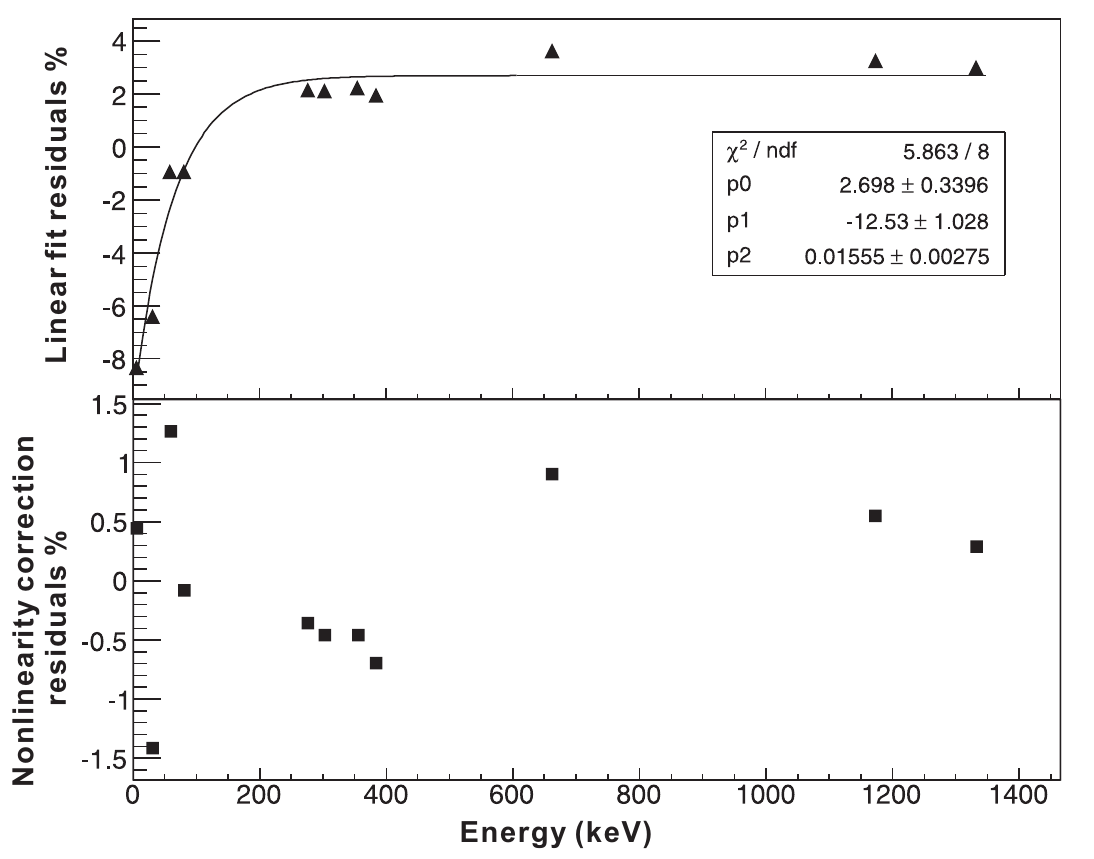}
\caption{\label{E-C} Left top: Energy to ADC channel conversion. Energy versus normalized peak channel position are plotted with linear fit. Left bottom: peak channel position divided by energy is plotted against energy. Right: Comparison between linear fit residuals (top) and nonlinearity correction residuals (bottom). Linearity distortion mainly occurs at energy below 300 keV. The nonlinearity correction reduces residuals.}
\end{figure}

 At first, the ADC data were subtracted by pedestals of collected pulse height histograms in Section 3.2. Then the ADC data were normalized according to the gain of the low gain and high gain channel.
To measure the energy to channel conversion (E-C conversion), the full energy peaks of the normalized pulse height histograms were fitted with Gaussian functions in order to extract the peak centriods and FWHM. In Fig.\ref{E-C} left top, the peak channel data versus their nominal energies were plotted with linear fit. The relation between the ADC channel (Chn) and energy is carried out using a linear function with only one coefficient, as
\begin{equation}
Chn=1.921\times E\label{Raw-EC}
\end{equation}
\par To more intuitively display the non-linearity, ADC peak channel divided by the energy is plotted against energy (Fig.\ref{E-C} left bottom). The value should be constant over the whole energy range for a good linear energy response; unfortunately, {\color{black}we observed an obvious distortion of linearity below 300 keV (Fig.\ref{E-C} left bottom) and there is a obvious channel/energy value decrease in the lower energy range\cite{NPR}}. {\color{black} The nonlinearity description is carried out using the residual as:}
\begin{equation}
Residual (\%)=100\times \frac{Observed-Expected}{Expected}\label{NPR_def}
\end{equation}
\par The linear fit residuals were plotted in Fig\ref{E-C} right top. The residuals can be fitted by a proposed evaluation function:
\begin{equation}
Res_{L}(\%) =p_{0}+p_{1} \cdot e^{-p_{2}\cdot E}\label{NPR_res}
\end{equation}
\par The fit parameters are \emph{p$_{0}$=2.698}, \emph{p$_{1}$=-12.53},  \emph{p$_{2}$=0.01555} and the $\chi$$^{2}$/ndf is 5.863/8. The energy to channel conversion is not proportional and nonlinearity correction is needed. The \emph{Chn} is corrected by:
\begin{equation}
Chn_{corr}=\frac{Chn}{1+Res_{L}(\%)/100}\label{NPR_correction}
\end{equation}
\par The residuals of the corrected data, as shown in Fig.\ref{E-C}right bottom, are within 1.5\%. The proposed evaluation function (Eq.\ref{NPR_res}) yet requires more precise energy calibration tests in the energy range below 300 keV.
\subsection{In-flight calibration capability with LaBr$_{3}$:Ce intrinsic gamma-ray lines}
\par
\begin{figure}[htbp]
\centering
\includegraphics[width=8cm]{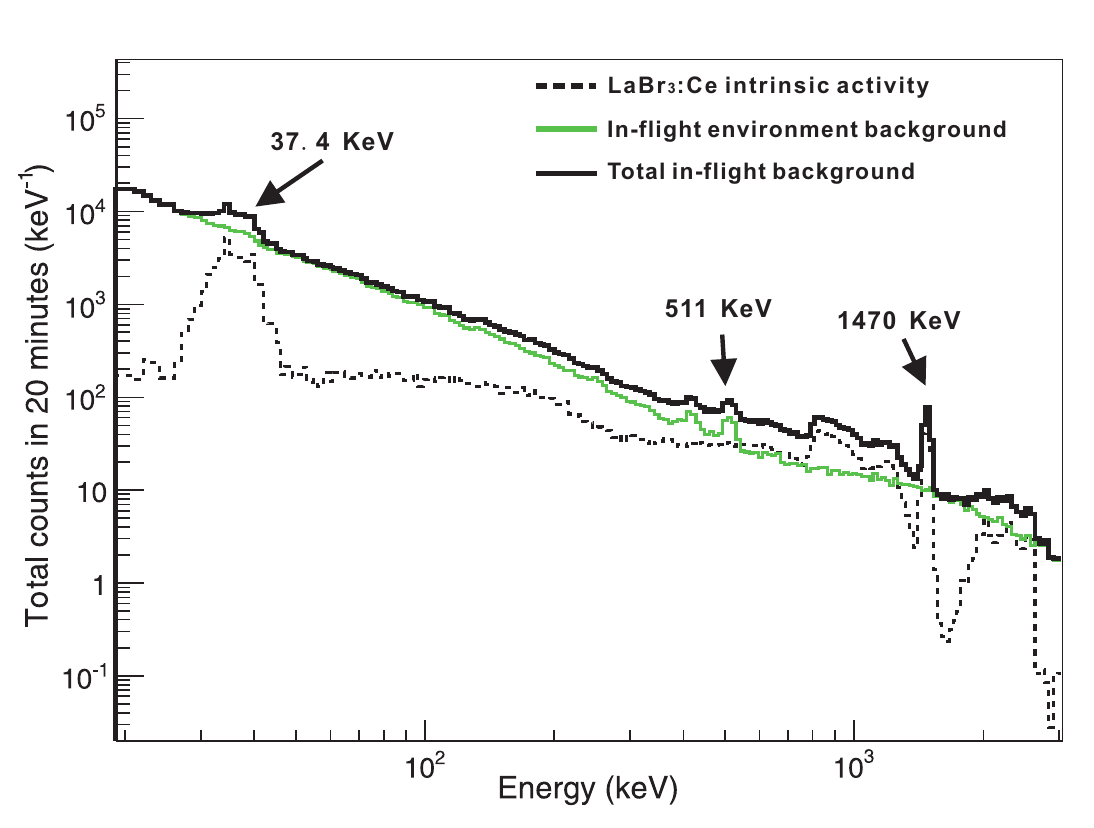}
\caption{\label{InflightCal} Calibration capability with LaBr$_{3}$:Ce gamma-ray lines. Black dotted curve is measured LaBr$_{3}$:Ce intrinsic activity, green curve is simulated in-flight environment background, back curve is total in-flight background. The 37.4 keV and 1470 keV intrinsic gamma-ray lines of LaBr$_{3}$:Ce can be resolved from in-flight backgrounds. The galactic 511 keV gamma-ray line also can be resolved.}
\end{figure}

 Drift in the detector energy response and readout electronics gain are common problems in space experiments. The payload works for several years and the state of the whole detector system may change. In-flight energy calibration is usually performed on scientific satellites by accumulating a spectrum with radioactive sources\cite{Inflight_cal} or the gamma-ray lines produced by materials in the spacecraft structure\cite{Struct_cal}. The galactic 511 keV gamma-ray line\cite{511keV_cal} can also be used for in-flight calibration. The 37.4 keV and 1470 keV intrinsic gamma-ray lines of LaBr$_{3}$:Ce provide the in-flight calibration of the GRD energy response. To evaluate whether the two intrinsic gamma-ray lines can be used for the in-flight calibration, we performed Geant4 based Monte Carlo simulation of GECAM GRD all mass model. The main in-flight environment background (cosmic X-ray background, SAA proton activated, albedo gamma, and cosmic proton)\cite{Inflight_BK} were added with an energy spread using Eq.(\ref{Res_Fit}); the results were shown in Fig.\ref{InflightCal}. 
In Fig.\ref{InflightCal}, the two intrinsic gamma-ray lines of LaBr$_{3}$:Ce can be resolved from the in-flight backgrounds (LaBr$_{3}$:Ce intrinsic activity and in-flight environment background). These results show the GRD based on SiPM array can use the LaBr$_{3}$:Ce intrinsic activity for in-flight calibration.

\section{Conclusions and future work}
\par A novel GECAM GRD module based on LaBr$_{3}$:Ce and SiPM Array was proposed. In this study, a set of radioactive sources tests was performed on the GECAM GRD module based on LaBr$_{3}$:Ce and SiPM Array. The radioactive source tests cover the energy range of 5.9--1332.5 keV and the tested energy resolution of 5.3\% at 662 keV FWHM meets GECAM requirements (8\%). The energy to channel conversion shows obvious nonlinearity in the energy range below 300 keV. A nonlinearity correction was performed to reduce the residuals and the residuals are within 1.5\%. We also investigated the feasibility of using LaBr$_{3}$:Ce intrinsic activity as an in-flight calibration method. The Geant4 based in-flight environment background simulation and measured GRD LaBr$_{3}$:Ce intrinsic activity were used to evaluate the capability of in-flight calibration. The two gamma-ray lines (37.4 keV and 1470 keV) of LaBr$_{3}$:Ce were resolved from the total in-flight background.

\par The GRD performance will be further optimized in the following GECAM development phase. An important work of further developments is to design a customized circle SiPM array rather than the currently used square SiPM array. It will have a better uniformity and energy resolution in the detection of high energy gamma-rays. The GRD non-linearity in the low energy range also needs to be further investigated and experiments were plan to carry out at the X-ray calibration facilities built by HXMT of IHEP and National Institute of Metrology in China.

\par {\color{black}The radiation damage of the LaBr$_{3}$:Ce crystal and SiPM array may push the low energy limit for gamma-ray detection well above the required 8 keV. In the next development phase,the total dose radiation tests on LaBr$_{3}$:Ce crystal and SiPM array will be carried out to get an evaluation of the performance degradation according to the in-flight radiation dose estimation.}

{\color{black}}
\section*{\itshape{{\color{black} Acknowledgements}}}
\par {\itshape {\color{black}We would like to express our appreciation to the staff of the Key Laboratory of Particle Astrophysics who offer great help in the phase of development. This research was Supported by Key Research Program of Frontier Sciences, Chinese Academy of Sciences, Grant NO. QYZDB-SSW-SLH012. The authors also would thank the anonymous reviewers for their detailed and constructive comments in evaluation this paper.}}

\bibliographystyle{elsarticle-num}



\end{document}